\documentclass[12pt]{article}
\usepackage{CJK}
\usepackage{graphicx}
\usepackage{float}
\usepackage{amscd, amssymb, mathtools}
\usepackage{array}
\usepackage{enumerate}
\usepackage{float}
\restylefloat{table}
\usepackage{hyperref}
\usepackage{authblk}

\author[1]{Anthony Bordg}
\author[2]{Yijun He}
\affil[1]{Department of Computer Science and Technology, University of Cambridge}
\affil[2]{University of Cambridge}

\title{Comment on ``Quantum Games and Quantum Strategies''}
\date{}

\begin{document}
	\maketitle

\begin{abstract}
	We point out a flaw in the unfair case of the quantum Prisoner's Dilemma as introduced in the pioneering Letter {\em Quantum Games and Quantum Strategies} of Eisert, Wilkens and Lewenstein. It is not true that the so-called {\em miracle move} therein always gives quantum Alice a large reward against classical Bob and outperforms {\em tit-for-tat} in an iterated game. Indeed, we introduce a new classical strategy that becomes Bob's dominant strategy, should Alice play the miracle move. Finally, we briefly survey the subsequent literature and turn to the 3-parameter strategic space instead of the 2-parameter one of Eisert {\em et al}.
\end{abstract}

Along with Meyer \cite{Meyer99} Eisert, Wilkens and Lewenstein pioneered the field of quantum game theory. In their classic Letter \cite{EWL} they investigated the quantization of the Prisoner's Dilemma. However, as previously noted \cite{CommentEWL} their restricted strategic space using two parameters
\[
\hat{U}(\theta,\phi)=
\begin{pmatrix}
e^{i\phi}\cos(\theta/2) & \sin(\theta/2) \\
-\sin(\theta/2) & e^{-i\phi}\cos(\theta/2)	
\end{pmatrix}
\]
is only a subset of $SU(2)$ and as a consequence is unlikely to reflect any reasonable physical constraint. Fortunately, this subset exhibits interesting properties arising in the quantum regime.	But below we show their section on the quantum-classical version of the Prisoner's Dilemma, where Alice may use a quantum strategy while Bob is restricted to a classical strategy, is also flawed. \\
In particular the claim that the so-called miracle move $\hat{M}\coloneqq \hat{U}(\pi/2,\pi/2)$ gives Alice ``at least reward $r = 3$ as pay-off, since $\$_A(\hat{M},\hat{U}(\theta,0)) \geq 3$ for any $\theta\in [0,\pi]$, leaving Bob with $\$_B(\hat{M},\hat{U}(\theta,0)) \leq \frac{1}{2}$'' \cite[p.3079]{EWL} is false. Indeed, for a maximally entangled game $\gamma = \frac{\pi}{2}$, for $\theta = \frac{\pi}{2}$ one has 
\[
\frac{1}{2} < \$_A(\hat{M},\hat{U}(\frac{\pi}{2},0)) = \$_B(\hat{M},\hat{U}(\frac{\pi}{2},0)) = 1 < 3\;.
\]
In the situation where Alice plays the miracle move while Bob is restricted only to classical strategies, for $0 \leq \gamma \leq \frac{\pi}{2}$ we have
\begin{align}
\$_A(\hat{M},\hat{U}(\theta,0)) & = \frac{1}{8}\, (21 + \cos(\gamma)^{2} (-3 + 14 \cos\theta) + 3 \sin(\gamma)^{2}-16 \sin\gamma\sin\theta) \\
\$_B(\hat{M},\hat{U}(\theta,0))  & = \frac{1}{8}\, (11 + \cos(\gamma)^{2} (7-6 \cos\theta)-7 \sin(\gamma)^{2} + 4 \sin\gamma\sin\theta)\;.
\end{align}
So, pluging $\gamma = \frac{\pi}{2}$ in equations (1) and (2) gives 
\[
\$_A(\hat{M},\hat{U}(\theta,0)) - \$_B(\hat{M},\hat{U}(\theta,0)) = \frac{5}{2}(1 -\sin\theta)
\]
admitting a minimum of $0$ when $\theta=\frac{\pi}{2}$. \\
In other words, in the 2-parameter scheme there is no miracle move and the dilemma is not removed in favor of the quantum player contrary to the claim in \cite[III.C]{FAIntro} which reproduced the faulty analysis of \cite{EWL} supported by erroneous computations (the authors found $\$_A = 3 + 2\sin\theta$ and $\$_B = \frac{1}{2}(1 - \sin\theta)$ instead of $\$_A = 3 - 2\sin\theta$ and $\$_B = \frac{1}{2}(1 + \sin\theta)$).  \\
Indeed, Bob can immunize himself against Alice's miracle move by playing the {\em down-to-earth} move $\hat{E}$
\[
\hat{E}\equiv\hat{U}(\frac{\pi}{2},0) = \frac{1}{\sqrt{2}}
\begin{pmatrix}
	1 & 1 \\
	-1 & 1
\end{pmatrix}\;,
\]
the outcome being a draw $\$_A = \$_B = 1$. \\
Following the notations in \cite{EWL} and assuming $\gamma=\frac{\pi}{2}$, $\phi_B = 0$, we get the following pay-off matrix.

\begin{table}[H]
	\centering
	\setlength{\extrarowheight}{2pt}
	\begin{tabular}{cc|c|c|c|}
		& \multicolumn{1}{c}{} & \multicolumn{3}{c}{Bob} \\
		& \multicolumn{1}{c}{} & \multicolumn{1}{c}{$\hat{C}$}  & \multicolumn{1}{c}{$\hat{D}$}  & \multicolumn{1}{c}{$\hat{E}$} \\\cline{3-5}
		& $\hat{C}$ & $(3,3)$ & $(0,5)$ & $(\frac{3}{2},4)$ \\ \cline{3-5}
		Alice  & $\hat{D}$ & $(5,0)$ & $(1,1)$ & $(3,\frac{1}{2})$ \\ \cline{3-5}
		& $\hat{Q}$ & $(1,1)$ & $(5,0)$ & $(3,\frac{1}{2})$ \\ \cline{3-5}
		& $\hat{M}$ & $(3,\frac{1}{2})$ & $(3,\frac{1}{2})$ &$(1,1)$ \\ \cline{3-5}
	\end{tabular}
\end{table}
So, if Alice plays $\hat{M}$, the dominant strategy of Bob becomes $\hat{E}$, thereby doing substantially worse than if they would both cooperate, reproducing the dilemma. Moreover, nothing supports the claim that Alice ``may choose ``Always-$\hat{M}$'' as her preferred strategy in an iterated game. This certainly outperforms {\em tit-for-tat} [\dots]'' \cite[p.3079]{EWL}. \\
In conclusion, the ``miracle move'' is of no advantage and there is nothing special about it. \\
We now turn to the 3-parameter scheme as outlined in \cite[3]{10.2307/3560138} for a brief comparison with the 2-parameter case. A pure quantum strategy becomes any $SU(2)$ operator
\[
\hat{U}(\theta, \alpha, \beta) =
\begin{pmatrix}
e^{i\alpha}\cos(\theta/2) & ie^{i\beta}\sin(\theta/2) \\
ie^{-i\beta}\sin(\theta/2) & e^{-i\alpha}\cos(\theta/2)
\end{pmatrix}\;,
\]
where $\theta\in [0,\pi]$ and $\alpha, \beta\in [-\pi,\pi]$. In the maximally entangled case $\gamma = \frac{\pi}{2}$ with Bob restricted to classical strategies ($\alpha = \beta = 0$), we get the following payoff matrix.

\begin{table}[H]
	\centering
	\setlength{\extrarowheight}{2pt}
	\begin{tabular}{cc|c|c|c|}
		& \multicolumn{1}{c}{} & \multicolumn{3}{c}{Bob} \\
		& \multicolumn{1}{c}{} & \multicolumn{1}{c}{$\hat{C}$}  & \multicolumn{1}{c}{$\hat{D}$}  & \multicolumn{1}{c}{$\hat{E}$} \\\cline{3-5}
		& $\hat{C}$ & $(3,3)$ & $(0,5)$ & $(\frac{3}{2},4)$ \\ \cline{3-5}
		Alice  & $\hat{D}$ & $(5,0)$ & $(1,1)$ & $(3,\frac{1}{2})$ \\ \cline{3-5}
		& $\hat{M}$ & $(\frac{3}{2},4)$ & $(\frac{3}{2},4)$ &$(3,3)$ \\ \cline{3-5}
	\end{tabular}
\end{table}

The miracle move guarantees Alice a minimum payoff against Bob's classical strategies, but the classical player is the one who benefits most from this move! Strangely enough, this point is not mentioned in \cite{10.2307/3560138}.

\section*{Acknowledgments}

This work was supported by the European Research Council Advanced Grant ALEXANDRIA (Project 742178).

\end{document}